\documentclass[pre,aps,floatfix,nofootinbib,superscriptaddress,two column]{revtex4-2}
\usepackage{graphicx}
\usepackage{dcolumn}
\usepackage{latexsym}
\usepackage{hyperref}
\usepackage{amsmath, amsthm, amssymb}
\usepackage{epsfig}
\usepackage{bm}
\usepackage{geometry}
\usepackage[dvipsnames]{xcolor}
\usepackage{placeins}
\usepackage{diagbox}
\begin{document}

\preprint{APS/123-QED}

\title{Competing effect of disorder on phase separation in active systems}

\author{Pratikshya Jena$^a$}
\email[]{pratikshyajena.rs.phy20@itbhu.ac.in} 
\affiliation{Indian Institute of Technology (BHU), Varanasi, 221005, India}

\author{Shambhavi Dikshit$^a$}
\email[]{sham_29@nus.edu.sg; Current affiliation: Mechanobiology Institute, National University of Singapore, 117411 Singapore}
\affiliation{Indian Institute of Technology (BHU), Varanasi, 221005, India}
\author{Shradha Mishra}
\affiliation{Indian Institute of Technology (BHU), Varanasi, 221005, India}

\date{\today}

\begin{abstract}
 We investigate the impact of  random pinned disorder on a collection of self-propelled particles. To achieve this, we construct a continuum model by formulating the coupled hydrodynamic equations for slow variables: local density and momentum density of particles. The disorder in the system acts as pinning sites, effectively immobilizing the particles that come into contact with them. Our numerical results reveal that weak disorder leads to phase separation in the system at density and activity lower than the typical values for motility induced phase separation. We construct a phase diagram using numerical simulations as well as linearized approximation in the plane of activity and packing fraction of particles at weak disorder densities. On increasing  disorder density the system shows the Micro phase separation, while at large disorder densities, the system becomes heterogeneous and eventually undergoes kinetic arrest.
 The structure factor tail deviates from the Porod's law, indicating increased roughness at domain interfaces under strong disorder. Furthermore, we analyze the fractal dimension of the interface as a function of disorder density, highlighting the increasing irregularity of phase-separated domains. We also found that disorder significantly suppresses number fluctuations in the system.
   
\end{abstract}

\maketitle


\section{\label{sec:level1}Introduction}
Nonequilibrium living and lab designed active systems for example, cells, tissues, living organisms and autonomous robots \cite{ramaswamy2010mechanics,ramaswamy2017active,marchetti2013hydrodynamics,vrugt2024review,volpe2022active} are composed of many self-propelled agents unveil intriguing collective behavior  across a wide range of length and time scales. The study of active matter systems became an interesting area of research for many years due to the emergent behaviors such as pattern formation \cite{denk2020pattern,hallatschek2023proliferating,jena2024spatio}, nonequilibrium disorder to order transitions \cite{KUMAR2024129773}, anomalous fluctuations \cite{jena2023ordering,ramaswamy2003active}, and interesting behavior in different medium and confinement etc., that are not present in corresponding equilibrium systems. Another captivating characteristic of such systems is Motility Induced Phase Separation (MIPS)  that resembles the passive liquid-gas phase separation but occurs in absence of any attractive interactions \cite{fily2012athermal,cates2015motility}, \cite{PhysRevX.9.031043,omar2023mechanical} at much lower packing densities. \\

The majority of studies of active matter in theory and experiments are focused on the systems in homogeneous or clean environments \cite{julicher2018hydrodynamic,sanchez2012spontaneous,zottl2016emergent,alston2022intermittent,aranson2022bacterial,gachelin2014collective,pattanayak2018collection,sampat2021polar,takatori2016forces,fily2012athermal,wittkowski2014scalar,stenhammar2013continuum,pattanayak2021ordering}.  But, in natural active matter systems, inhomogeneity or disorder is present intrinsically. Various types of disorder or inhomogeneities are observed in natural systems that arise from multiple factors. This can be present in the form of spatial (geometric variation), temporal (moving obstacles), chemical composition and, biological inhomogeneity (bio-diversity) etc. Exploring and understanding such systems with disorder is essential to realize the complexity of natural systems and the systems' response to internal as well as external stimuli.\\
There have been extensive studies on active systems in clean environments, while the exploration of systems with disorder is relatively less. However, there exist a few prior studies on systems possessing disorder \cite{kumar2020active,das2018polar,chen2022hydrodynamic,vahabli2023emergence,pincce2016disorder,ro2021disorder,goswami2022motion,mishrainhomogeneous,kumar2021effect}.
These studies explore the effects of both quenched and annealed disorder in active systems, highlighting their impact on the dynamics of active particles by examining transport properties and diffusivity.  In certain cases, disorder enhances collective behavior in active polar particles, while in active scalar particles, it has been reported to induce long-range order \cite{ro2021disorder}. The works by C. Reichhardt and C. J. O. Reichhardt have made significant contributions to the study of systems in disordered media over many years. Using microscopic methods, their work explores a variety of scenarios, including systems with periodic arrays of obstacles and inhomogeneous substrates \cite{canavello2025polarization,souza2024skyrmion,reichhardt2024stripe,minogue2024reversible}. Their findings \cite{canavello2025polarization,souza2024skyrmion} demonstrate that disorder can give rise to distinct dynamical phases. Notably, they observe phenomena such as pinning, ratchet effects, clogging, collective behavior, and pattern formation \cite{reichhardt2023pattern,libal2023dynamic,zhu2022directional} etc. \\
In most of the other studies, disorder acts as either obstacles placed physically in the space or a random field disorder \cite{das2018polar, chepizhko2013diffusion, ro2021disorder, morin2017distortion,das2018ordering}. 
Apart from some studies, very little is still known about the phase behavior of the active systems in the presence of disorder which acts like pinning sites and immobilize the particles. The pinning sites as a disorder can be more clearly understood from a biological point of view. Pinned or “frozen” sites correspond to heterogeneous environments where cells or active agents cannot move and get stuck due to the mechanical trapping. There are plenty of examples of such biological systems that naturally contain such immobile, obstructive patches. 
This kind of fixed obstacles can be noticed in many natural system i.e, motion of bacteria in porous media \cite{drescher2011fluid}, migrating cells encountering collagen fibers \cite{wolf2011extracellular}, collective cell migration through tissue environment \cite{petrie2009random} etc. \\
Our primary focus in the present work is on the effect of spatial heterogeneity, modeled as pinned disorder, on phase behavior, phase separation kinetics, and resulting morphology of active matter systems using a field theoretic approach, which sets it apart from previous works. The study will be beneficial for observing the dynamical behavior of the organism in the presence of disorder in natural systems. 

In the present work, we introduce disorder, as pinning sites in a system consisting of a collection of active particles,  by constructing a continuum model by formulating the coupled hydrodynamic equations for slow variables: local density and momentum density of particles.  The pinning sites are modeled in such a manner that they effectively make the self-propelled speed of the active particles zero. We explore the phase diagram to gain insight into the phenomenon of phase separation in the presence of disorder. We observed phase separation at density and activity lower than the critical values for a clean system \cite{fily2012athermal, redner2013structure}, a result further confirmed by linearized hydrodynamic calculations. Although, the pinned disorder eases phase separation, the phase separation weakens with increasing disorder density. At moderate disorder densities Micro phase separation is found whereas the arrested kinetics is found at high disorder densities. The structure factor tail exhibits a deviation from the Porod's law, suggesting enhanced roughness at domain interfaces in the presence of strong disorder. Additionally, our analysis of the fractal dimension of the interface as a function of disorder density reveals irregularity in phase-separated domains.\\

The structure of this paper has been organized as follows. Sec \ref{sec:model} provides the detail explanation of our model along with the numerical methodology for solving the equations. We demonstrate the outcomes of our research in sec \ref{sec:result}. In sec \ref{sec:discussion}, we discuss our interesting notable findings and present a summary highlighting the relevance of our study.


\section{Model and Numerical Methodology}\label{sec:model}
We develop a continuum model to study a collection of self-propelled particles on a two dimensional substrate with the presence of random pinning sites. We formulate a  minimal hydrodynamic coupled equations for conserved  local density field $\rho(\boldsymbol{r},t)$ and local orientation order parameter $\boldsymbol{p}(\boldsymbol{r},t)$ and further, we define the momentum density $\boldsymbol{P}(\boldsymbol{r},t) = \rho(\boldsymbol{r},t)\boldsymbol{p}(\boldsymbol{r},t)$.
 The hydrodynamic equation for the density $\rho(\boldsymbol{r},t)$ is:
 \begin{equation}
    \partial_{t}\rho= -\boldsymbol{\nabla}.(v(\rho)\boldsymbol{P}-D_{\rho}\boldsymbol{\nabla}\rho+\boldsymbol{f}_{\rho})
    \label{eq:r}
    \end{equation}
  
and for the local momentum  density $\boldsymbol{P}(\boldsymbol{r}, t)$ is:
   \begin{equation}
    \partial_{t}\boldsymbol{P}=-\nu_{r} \boldsymbol{P}-\frac{1}{2}\boldsymbol{\nabla}(v(\rho)\rho)+k\boldsymbol{\nabla}^{2}\boldsymbol{P}+\boldsymbol{f_{P}}
    \label{eq:v}
\end{equation}

Here, $v(\rho) = v_0(1-\lambda \rho)$, effective velocity of each particle, is influenced by the local density of the particles and the presence of pinned sites in the system. A pinned disorder site corresponds to a rectangular region of size $\Delta x\times \Delta x$ and these regions are chosen randomly initially and then their spatial arrangement is quenched with time.
The velocity dependence on the local density is adopted from previous observations that the local clustering suppresses the motile nature of particles \cite{fily2012athermal}. Further, we introduce the pinned obstacles or disorder in such a manner that there are few randomly selected regions in the substrate such that $v(\rho) = 0$ at those points, still the particle can diffuse because the noise $\boldsymbol{f}_{\rho}$ is not zero. Here, we mean the particle can only diffuse and lose its motility and thermal fluctuations can still lead them to diffuse. The number density of such points on the substrate is the density of pinning sites in the system, and is defined as $\rho_d$.  \\
 Eq.\ref{eq:r} is a continuity equation where, the first term on the right side of the equation accounts for the active current generated by self-propulsion  and the second term  represents the diffusion current with diffusion constant $D_\rho$. Eq. \ref{eq:v} is similar to the equation introduced by Fily et.al \cite{fily2012athermal} in the context of self-propelled particles neglecting the nonlinearites in $\boldsymbol{P}$. The first and second terms on the right hand side of Eq. \ref{eq:v} represent the polarization decay at rate $\nu_{r}$ and it's convection by pressure-like gradients $\sim \boldsymbol{\nabla}(v(\rho)\rho)$ respectively. The third term represents diffusion in orientation.
And $\boldsymbol{f}_{\rho}$ and $\boldsymbol{f}_{\boldsymbol{P}}$ are the Gaussian white noise in density and polarization equations respectively having strengths $\Delta_{\rho, \boldsymbol{P}}$;
\begin{equation*}
    \langle f_{\alpha,i}({\bf r},t)f_{\alpha, j}({\bf r^{'}},t{'}) \rangle = \Delta_{\alpha}\delta_{ij}\delta({\bf r}-{\bf r}^{'})\delta(t-t^{'})
\end{equation*}
\\
where, $\alpha \equiv (\rho, \boldsymbol{P})$ and $(i,j)$ can take two values $x$ and $y$ representing the two cartesian coordinates. The noise we have taken $\boldsymbol{f}_{\rho}$ and $\boldsymbol{f}_{\boldsymbol{P}}$ are additive in nature.  For simplification in the model, we take the noise as additive which gives random kick and does not depend on the local density.
The intrinsic time scale $\tau$ and intrinsic length scale $l_0$ are defined as $\nu_r^{-1}$ and $\sqrt{D_{\rho}/\nu_r}$ respectively. Eq. \ref{eq:r}, \ref{eq:v} are rescaled by the $\tau$ and $l_0$.
 Numerical integration of Eq. \ref{eq:r} and \ref{eq:v} is performed with homogeneous initial density with mean $\rho_0$ and random ${\boldsymbol{P}}$. We vary the mean density from $\rho_0$ = 0.2 to 0.9 and self-propulsion speed from $v_0 = 0$ to $7$ in a box of size $K \times K$ with periodic boundary condition in the both directions. To investigate the impact of disorder in the system, we vary the disorder density $\rho_d$ from $0$ to $0.5$.  The integration is performed using Euler’s scheme \cite{bally1996law} with $\Delta x = 1.0l_0$ and $\Delta t = 0.1\tau$.
 In the system under consideration, the parameters are fixed as follows: $\nu_{r}= k = D_{\rho}=1.0$ and  $\lambda=0.9$. For these choice of parameters, the initial homogeneous state becomes unstable and system phase separates for $\rho_0 > 1/(2 \lambda)$ as reported in \cite{fily2012athermal}.  We performed numerical simulations for simulation time, $t=2 \times 10^5$ and for  system sizes $K = 256 - 1024$. Further, the averaging is performed over  $50$ independent realizations to get better statistics. 

 \begin{figure*}
     \centering
     \includegraphics[width=0.8\textwidth]{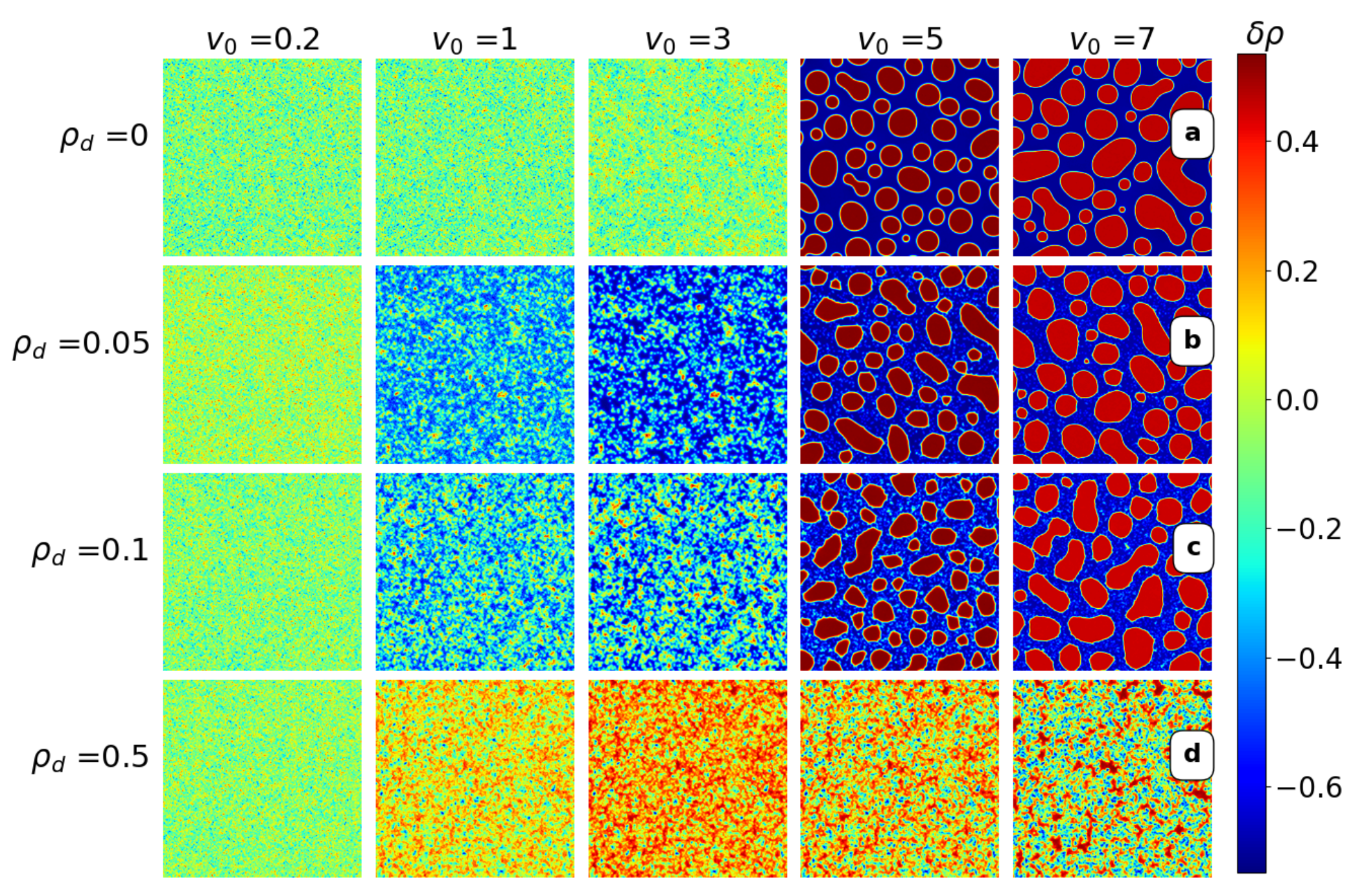}
     \caption{ The panels (a-d) showcase the local fluctuation $\delta\rho$ at across four panels , each representing system with different disorder densities $\rho_d = 0.0, 0.05, 0.1, 0.5$ respectively at $t=200$. Within each panel, multiple figures are plotted from left to right for self-propulsion speed $v_0 = 0.2, 1, 3, 5, 7$ keeping the mean density $\rho_0 = 0.7$ fixed.  The color on the heatmap represents the value of $\delta\rho$. The results are obtained for systems having $K = 256$.}
     \label{fig1}
 \end{figure*}

 \begin{figure*}
     \centering
     \includegraphics[width=0.76\textwidth]{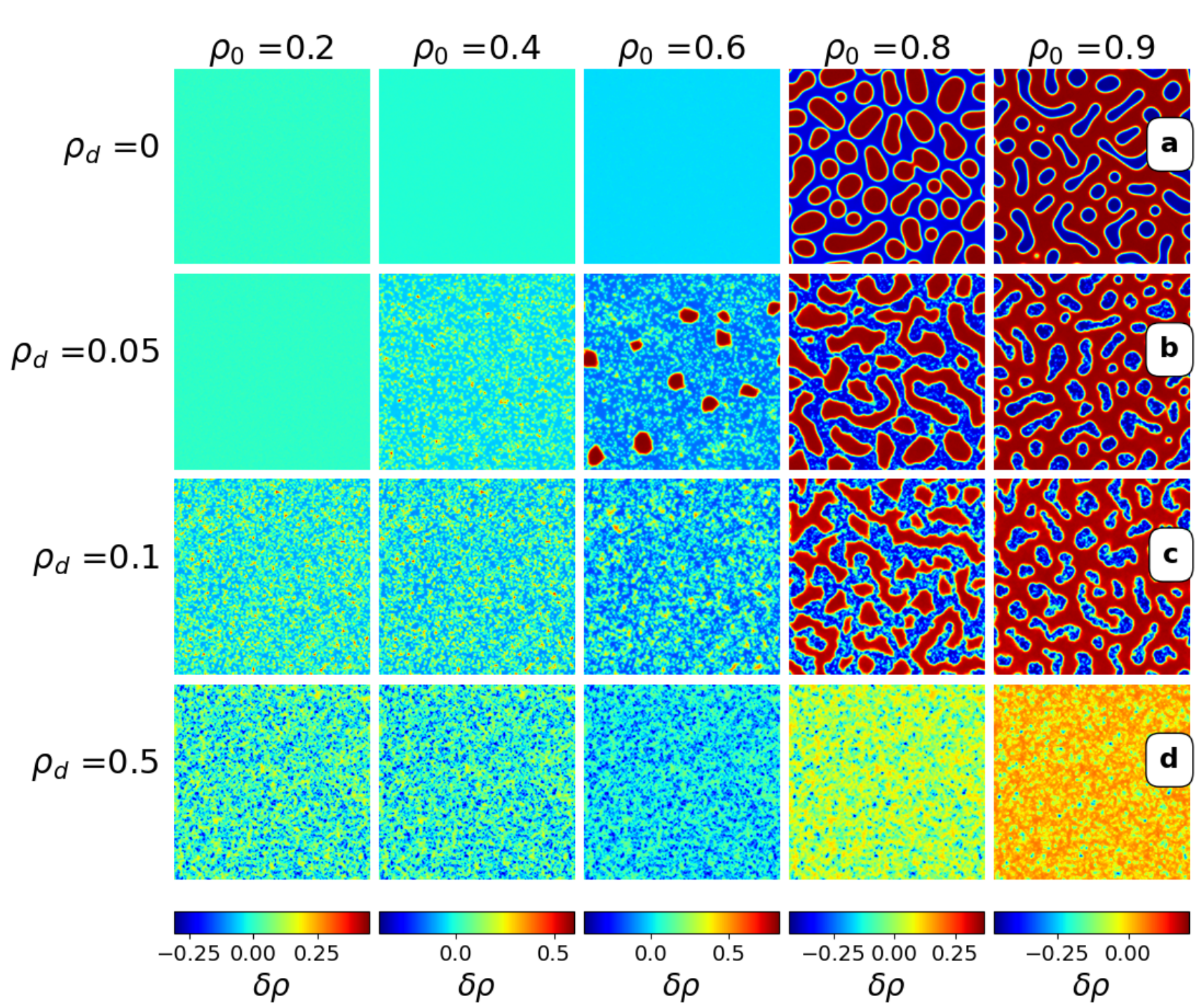}
     \caption{The panels (a-d) depict the local fluctuations in density, $\delta \rho$, at time $t = 200$ for systems with varying disorder densities, $\rho_d = 0, 0.05, 0.1, 0.5$ in sequence. Each panel consists of multiple snapshots, from left to right, representing systems with mean densities $\rho_0 = 0.2, 0.4, 0.6, 0.8, 0.9$ respectively by fixing the self-propulsion speed $v_0 = 4.0$. The color in the heatmap indicates the magnitude of the $\delta \rho$. The results are generated for systems having $K = 256$.}
     \label{fig2}
 \end{figure*}

 \begin{figure*}[hbt]
  \includegraphics[width=1.0\textwidth]{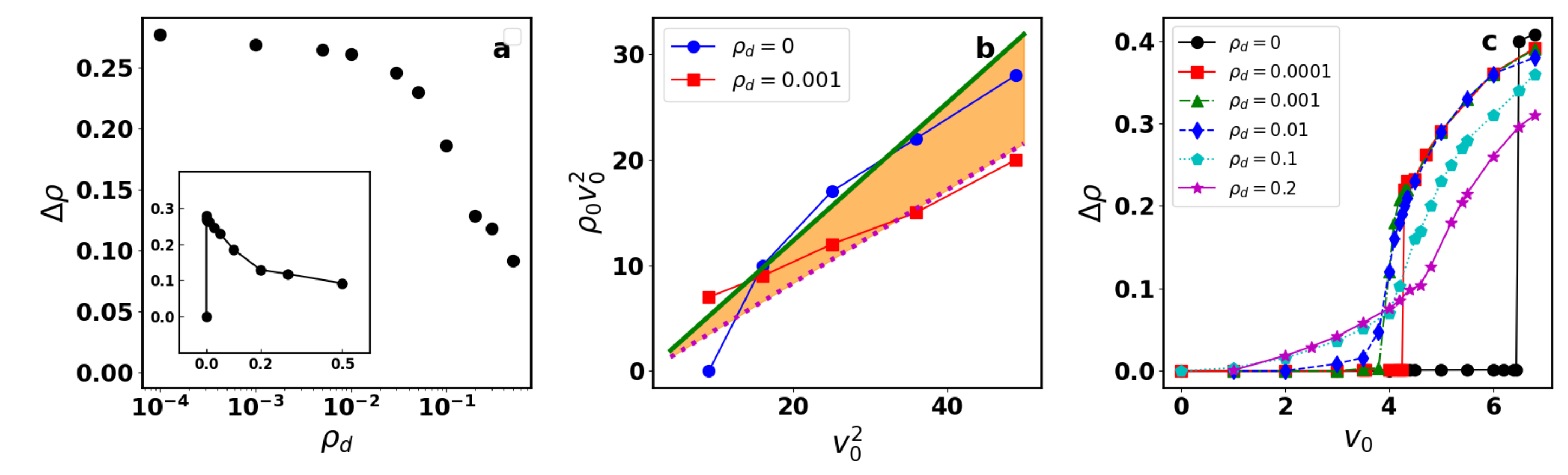}

  \caption{The plot(a) presents semi-log $x-$ plot of $\Delta\rho$  {\em vs.} $\rho_d$ with error bars. The inset shows a linear plot of $\Delta\rho$ {\em vs.} $\rho_d$ depicting the non-monotonicity with respect to disorder. 
  The plot(b) illustrates the phase diagram in the $\rho_0 v_0^2-v_{0}^2$ plane for different disorder densities: $\rho_d = 0.0, 0.001$ respectively. The blue (circles) and red (squares) solid lines are the boundary drawn from the numerical simulation for $\rho_d = 0.0, 0.001$ which is also mentioned in the legend. Additionally, the green solid line and magenta dotted line represent the analytical boundary obtained from linearized calculation. The shaded light-orange region shows the extra regime of phase separation in the presence of disorder.
The plot (c) showcases $\Delta\rho$  {\em vs.} $v_0$ depicting the transition from non-phase separation to phase separation for $\rho_d = 0.0-0.2$.  }
  \label{fig3}
\end{figure*}

\section{Results}\label{sec:result}
{\em{Snapshots for local density fluctuation}}: 

To investigate the effect of disorder on the system, we begin by plotting snapshots of normalized local density fluctuations (${\delta \rho}$) defined as $ {\delta \rho}({\boldsymbol{r}}, t) =\frac{\rho({\boldsymbol{r}}, t)-\rho_0}{\rho_0} $. We examine  the result for a range of ($v_0$, $\rho_0$) values in systems with varying density of disorder $\rho_d$. Figs. \ref{fig1}(a-d) illustrate snapshots of local density fluctuations across four panels, showcasing the effect of varying $\rho_d$ and fixing $\rho_0 = 0.7$. These snapshots are taken at fixed time $t = 200$ (simulation time $t=2000$) which does not correspond to a steady state, it is in a transient state.
Panel (a) represents a system without disorder ($\rho_d = 0$) for different values of $v_0 = 0.2, 1, 3, 5, 7$. 
 Panels (b), (c), and (d) display the corresponding snapshots for systems with disorder densities of $\rho_d$ = 0.05, 0.1, and 0.5, respectively. Fig. \ref{fig1} Panel (a) shows that in a clean system, phase separation begins at $v_0 > 3$ with the formation of small domains, indicating the onset of Motility-Induced Phase Separation $(MIPS)$, consistent with previous studies on active Brownian particles \cite{fily2012athermal}. 
 For the systems with non-zero disorder density, the local density fluctuations ${\delta \rho}$ shows non-zero values at $v_0 = 3$ as can be seen in Fig. \ref{fig1} Panel (b) for $\rho_d = 0.05$, indicating phase separation in the system. Thus, the density inhomogeneity or phase separation develops at a speed below that for  the clean system. As the numbers of pinning sites  increases, as shown in Fig \ref{fig1} Panel (c) for $\rho_d = 0.1$, the density inhomogeneity starts to appear at lower $v_0$. But, after a certain $\rho_d$ the system does not phase separate and instead shows the heterogeneous density as shown in Fig. \ref{fig1} Panel (d) for $\rho_d =  0.5$. \\
Similarly, in Fig. \ref{fig2} Panel (a-d), we show the fluctuations in local density for fixed $v_0 = 6$ , varying mean particle densities $\rho_0$ , for different disorder densities.
 The density of disorder is the same as for the Fig. \ref{fig1} Panel (a-d). This allows us to examine how both $\rho_0$ and disorder influence the system's behavior.
We observed that the system undergoes phase separation at ${\rho_0} > 0.6$ 
for the clean system ($\rho_d = 0$) in Fig. \ref{fig2} Panel (a). As suggested by previous studies for clean systems \cite{fily2012athermal}, $MIPS$ occurs once a critical density is reached.
However, as the disorder in the system increases, phase separation occurs even at lower values of $\rho_0$ as shown in Fig. \ref{fig2} Panel (b-c). Therefore, the diagram demonstrates that disorder promotes density inhomogeneity in the system for weak disorder. Similar to Fig. \ref{fig1} Panel (d), higher disorder makes system heterogeneous again as shown in Fig.\ref{fig2} Panel(d). \\
Based on the observation from the snapshots shown in Figs. \ref{fig1} and \ref{fig2}, it can be concluded that   systems with finite $\rho_d$ shows the phase separating domains at self propulsion speed $v_0$ and mean density $\rho_0$ smaller than that for the clean system $\rho_d = 0$. Hence, the presence of small number of  pinning sites drives the phase separation at lower activity and mean density. Further, the system with high density of disorder  shows the formation of heterogeneous  structures suppressing phase separation. 

\begin{figure*}
    \includegraphics[width=1.0\textwidth]{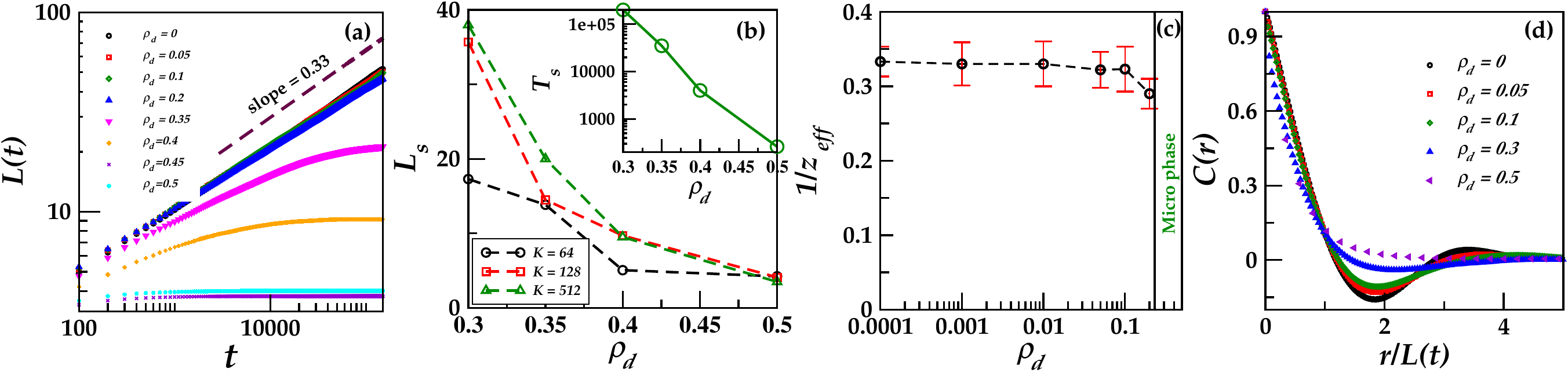}
    
    \caption{ The plot (a) showcases the log-log plot of characteristic length $L(t)$ {\em vs.} $t$ for different disorder density $\rho_d$ ranging from $0-0.5$ in the system showing in the legends. The dashed orange line depicts the line of slope $0.33$. The plot (b) shows the saturation length $L_s$ $vs.$ $\rho_d$ for different system sizes $K=64,128,512$. The inset depicts the semi-log $x-$ plot for saturation time $T_s$ $vs.$ $\rho_d$. The lines are guide to eyes, both in the main and inset panel. 
    The plot (c) displays semi-log $x-$ plot of the corresponding $1/z_{eff}$ with $\rho_d$. The error bars represent the standard deviations of the $1/z_{eff}$.
    The vertical line drawn here at the critical value of $\rho_d = 0.23$ (from Eq. \ref{critical_rho}) after which the phase separation is suppressed and is obtained from the linearized calculation. 
    The plot (d) presents the static scaled correlation $g(x)$ for $\rho_d = 0 -0.5$.}
    \label{fig6}
\end{figure*}

To analyze it thoroughly, in Fig. \ref{fig3}(a) we show the plot of density phase separation order parameter $\Delta\rho$ $(PSOP)$  {\em vs.} $\rho_d$, where 
$\Delta \rho = \frac{1}{K^2}<{\sum}_{\boldsymbol{r}} \vert \delta \rho(\boldsymbol{r}, t) \vert >$,   where $<..>$ mean average over time in the steady state and over different realizations. 
We calculated the $PSOP$ with respect to time and we define the steady state when the $PSOP$ does not significantly change with respect to time.
Here, $\Delta \rho$  reflects the amount of phase separation in the system.  We fix the  self-propelled speed $v_0$ and $\rho_0$ to values  4.2 and 0.7 respectively, such that the homogeneous state is stable so that $\Delta \rho = 0 $ for the clean system. As we introduce disorder, $\Delta \rho$ shows a jump to finite value and again decreases for the large disorder densities monotonically as shown in Fig. \ref{fig3}(a)(inset). A very small disorder is enough to make the homogeneous state unstable as shown in Fig. \ref{fig3}(a)(main panel) (on semi-$\log x$- scale). \\
We now investigate the mechanism responsible for disorder-induced phase separation at low disorder densities and the transition to a heterogeneous phase at high disorder densities.
In the low disorder environment, the accumulation around the obstacles could be a result of particles being immobilized upon reaching close proximity, which leads to effectively zero velocity causing the particles to gather at that location. Additionally, when new
particles approach, the existing ones act as obstacles and contributing to the reduction of their velocity. As a
result of this effect, clustering around the obstacles and domains formation is observed. Since, the phase separation is driven by the disorder in the system, we refer to it as Disorder-Induced Phase Separation ($DIPS$). The movie \href{https://drive.google.com/file/d/1DFcUoYSQ5x2F0yMM5zccSz8v0i4wCeF0/view?usp=sharing}{MV1} (MV1) shows how the pinning sites (represented as black dots) act as nucleation sites and accelerate the formation of cluster of the active particles around it. The color bar in the movie depicts the local density field $\rho(r)$ of the active particles.  To further quantify it, we have also calculated the correlation between active particles and disorder sites $C_{\rho \rho_d}(t)$. The mathematical definition of $C_{\rho \rho_d}$ is given in Appendix \ref{cor_t} It is a measure of likelihood of finding a cluster of particle near a disorder site. The positive $C_{\rho \rho_d}$ means clustering of particle close to disorder sites, or low density of particle, where there is no disorder. The plot is generated for low disorder density $\rho_d = 0.05$ as shown in Fig \ref{Fig:A2}. From the plot, the correlation is seen to start with a small positive value, then gradually increase, and eventually saturate at late times. This behavior indicates that, initially, clusters nucleate in the vicinity of the disorder sites. After sufficient time, the saturation of the correlation suggests that $MIPS$ becomes the dominant mechanism. We discussed this in details while explaining Fig. \ref{fig6} (a). 
However, for high $\rho_d$, the particles get immobilized at most of the sites. There is no scope for movement of the particles and they get stuck immediately around their nearest pinning sites.   As the particles immobilize immediately, they can not accumulate in sufficient amount to form large clusters. Although, the diffusion in density term is non-zero, but it simply diffuses the particles from one place to another and pinned sites act like cold regions with small accumulation of density around that sites. As a result,  heterogeneous structures are formed in the system. \\
{\em {Phase Diagram}}: 

To better illustrate the shift in the phase boundary, we present a phase diagram in Fig. \ref{fig3}(b). To gain further insight from the hydrodynamic equations, we perform a linearized analysis around the homogeneous state, demonstrating that in the activity ($v_0$) and mean density ($\rho_o$) plane, the phase boundary shifts towards lower values of both. The details of the calculation from the linearised hydrodynamic is given below.\\ 

\onecolumngrid
\noindent\rule{\textwidth}{0.8pt}

\begin{center}
{\em Linearise calculation: Shifting of phase boundary }
\end{center} 
 We develop a linearised calculation of the hydrodynamic equations provided in the model section and  show the shifting of phase boundary in the plane of ($\rho_0$ and $v_0$). 
The homogeneous steady state solution of the above equations are $\boldsymbol{P}$=$0$ and $\rho$=$\rho_{0}$. Adding small fluctuations to the steady state solutions i.e;
 $\boldsymbol{P}=\delta \boldsymbol{P}$ and $
      \rho= \rho_{0}+\delta{\rho}$
 the effective self-propulsion speed 
$  v_e(\rho, \rho_d) = v_0(1-\lambda\rho)A(r)$, where $\langle A(r)\rangle = 1-\rho_d$, and the mean $\langle..\rangle$ is over different realizations. $\langle A(r)\rangle = 1$ for the clean system. Further substituting the density in terms of small fluctuation $\delta \rho$,
$v_e = v_0(1-\lambda \rho_0) \langle A(r) \rangle-v_0 \lambda \delta\rho \langle A(r) \rangle $. 
Now, we define $V = v_0(1-\lambda\rho_0)\langle A(r) \rangle$.
Equation \ref{eq:r} and \ref{eq:v} becomes, 
\begin{equation}
    \partial_{t}\delta \rho = -\boldsymbol{\nabla} . V \delta \boldsymbol{P}+ D_{\rho} \boldsymbol{\nabla}^{2} \delta \rho- \boldsymbol{\nabla}.\boldsymbol{f_{\rho}}
    \label{A1}
\end{equation}
and 
 \begin{equation}
    \partial_{t} \delta \boldsymbol{P}= -\nu_{r} \delta\boldsymbol{P} -\frac{1}{2}V\delta\rho+\frac{1}{2}v_0\lambda\rho_0 \langle A(r) \rangle\nabla\delta\rho+k \boldsymbol{\nabla}^{2} \delta \boldsymbol{P} +\boldsymbol{f_{P}}
\end{equation}
Taking divergence of equation(5) and defining $\nabla. \delta\boldsymbol{P} = \theta$

\begin{equation}
    \partial_{t} \theta =-\nu_{r}\theta-\frac{1}{2}V\nabla^2\delta\rho+\frac{1}{2}v_0\lambda\rho_0\langle A(r) \rangle\nabla^2\delta\rho+k \boldsymbol{\nabla}^{2} \theta + \boldsymbol{\nabla}. \boldsymbol{f_{P}}
\end{equation}

Equation \ref{A1} becomes,
\begin{equation}
    \partial_{t}\delta\rho=-V\theta +D_{\rho}\boldsymbol{\nabla}^{2}\delta\rho- \boldsymbol{\nabla}. \boldsymbol{f}_{\rho}
\end{equation} 
Later we replace the notation $\langle A(r) \rangle$ by $A$ for simplicity.
To perform the mode analysis we write the Linearized equations for order-parameter  and for density  in Fourier mode,
\begin{equation*}
\theta(\boldsymbol{r},t)=  \int_{\boldsymbol{q},\omega} \Theta(\boldsymbol{q},\omega) \exp(-i\omega t+ i\boldsymbol{q}.\boldsymbol{r}) d\boldsymbol{q} d \omega
  \end{equation*}
\begin{equation*}
\delta\rho(\boldsymbol{r},t)=  \int_{\boldsymbol{q},\omega} \delta\rho(\boldsymbol{q},\omega) \exp(-i\omega t+ i\boldsymbol{q}.\boldsymbol{r}) d\boldsymbol{q} d \omega
  \end{equation*}
 The linear equations in Fourier mode are,
  \begin{equation}\label{Eq:l1}
      [-i\omega+D_{\rho}q^{2}] \delta\rho + {V}\Theta = -i \boldsymbol{q}.\boldsymbol{f}_{\rho}
  \end{equation}
  
  \begin{equation}\label{Eq:l2}
      [-\frac{1}{2}{V}+\frac{1}{2}v_0\rho_{0}\lambda A]q^{2} \delta\rho + [-i\omega+k q^{2}+\nu_{r}]\Theta=i \boldsymbol{q}. \boldsymbol{f_{P}}
  \end{equation}
 
 The linear equations can be easily solved for $\Theta$
  and     $\delta\rho$ using Matrix method. This can be written in 2 $\times$ 2 matrix form.
   \[
  \begin{bmatrix}
  -i\omega+D_{\rho}q^{2}            &                 {V} \\
  -\Omega q^2                           &               -i\omega+kq^2+\nu_{r}
 \end{bmatrix}
  \begin{bmatrix}
    \delta\rho \\ \Theta
  \end{bmatrix}
    = \begin{bmatrix}
    -i \boldsymbol{q}. \boldsymbol{f_{\rho}}  \\ i \boldsymbol{q}. \boldsymbol{f_{{P}}}
  \end{bmatrix}
 \]
 where $\Omega= \frac{1}{2} [V-v_{0}\rho_0\lambda  A]$. The solutions for $\Theta$ and $\delta\rho$ is;

 \begin{equation}
  \begin{bmatrix}
 \delta\rho  \\ \Theta

 \end{bmatrix}
 =  
 M^{-1}
 \begin{bmatrix}
   -i q f_{\rho}  \\ i q f_{p}
 \end{bmatrix}
 \label{A7}
  \end{equation}

 Where the matrix M is,
\[
\begin{bmatrix}
 M
 \end{bmatrix}
  =
  \begin{bmatrix}
  -i\omega+D_{\rho}q^{2}            &                 V \\
  -\Omega q^2                           &               -i\omega+kq^2+\nu_{r}
 \end{bmatrix}
 \]
  The inverse of 2 $\times$ 2 matrix is 
 \[
 \begin{bmatrix}
 M^{-1}
 \end{bmatrix}
 = 
 \frac{1}{det(M)}
 \begin{bmatrix}
 -i\omega+kq^2+\nu_{r}              &           -V  \\
  \Omega q^2          &          -\iota\omega+D_{\rho}q^{2}  
   \end{bmatrix}
 \]
 where 
 \begin{equation*}
     det[M]= -\omega^{2}-i\omega[(kq^{2}+\nu_{r})+D_{\rho}q^{2}] +D_{\rho}q^{2}(kq^{2}+\nu_{r})+V\Omega q^{2}
 \end{equation*}
 
 The two modes obtained are 
 \begin{equation}
     w_{ 1}= -i\frac{1}{2}[(D_{\rho}+k)q^{2}+\nu_{r}] -i\frac{1}{2}[((\nu_{r}+kq^{2})-D_{\rho}q^{2})^{2}-4V\Omega q^{2}]^{1/2}
 \end{equation}
  \begin{equation}
     w_{ 2}= -i\frac{1}{2}[(D_{\rho}+k)q^{2}+\nu_{r}] +i\frac{1}{2}[((\nu_{r}+kq^{2})-D_{\rho}q^{2})^{2}-4V\Omega q^{2}]^{1/2}
 \end{equation}
 We can consider fluctuation as a wave that has the form $e^{-i\omega t}$. This can be written as $e^{-iRe(w)}.e^{Im(w)}$. For instability $Im(w)>0$.
 Below we determine the condition for instability in the mode $\omega_2$,
  \begin{equation*}
     ((\nu_{r}+kq^{2})-D_{\rho}q^{2})^{2}-4V\Omega q^{2} > 0
 \end{equation*}
 and
 \begin{equation*}
     \sqrt{((\nu_{r}+kq^{2})-D_{\rho}q^{2})^{2}-4V\Omega q^{2}} > (D_{\rho}+k)q^{2}+\nu_{r}
 \end{equation*}
  \begin{equation*}
     ((\nu_{r}+kq^{2})-D_{\rho}q^{2})^{2}-4V\Omega q^{2} > [(D_{\rho}+k)q^{2}+\nu_{r}]^2
 \end{equation*}
 \begin{equation*}
    (\nu_{r}+kq^{2})^2+(D_{\rho}q^{2})^2-2 D_{\rho}q^{2}(\nu_{r}+kq^{2})-4V\Omega q^2-(D_\rho+k)^2 q^4-\nu_{r}^2-2\nu_{r}q^2(D_\rho+k)>0
 \end{equation*}
\begin{equation*}
    -4q^2[D_\rho (kq^2+\nu_r)+V\Omega]>0
\end{equation*}
To satisfy this, $V\Omega<0$ and $|V\Omega|>D_\rho (kq^2+\nu_r)$.\\
 {\em case-1}
$V\Omega<0$
\begin{equation*}
    \frac{1}{2}[v_0^2A^2 (1-\lambda\rho_0)(1-2\lambda\rho_0)]<0
\end{equation*}
\begin{equation*}
    (1-\lambda\rho_0)(1-2\lambda\rho_0)<0
\end{equation*}
{\em case-2}
$|V\Omega|>D_\rho (kq^2+\nu_r)$

\begin{equation*}
    \frac{1}{2}[v_0^2A^2 (1-\lambda\rho_0)(1-2\lambda\rho_0)]>D_\rho (kq^2+\nu_r)
\end{equation*}
\begin{equation*}
    (2\lambda^2\rho_0^2-3\lambda\rho_0+1)>\frac{2 D_\rho (kq^2+\nu_r)}{v_0^2A^2}
\end{equation*}
for small q,
\begin{equation*}
    2\lambda^2\rho_0^2+1-3\lambda\rho_0-\frac{2 D_\rho \nu_r}{v_0^2A^2}<0
\end{equation*}
\begin{equation*}
   \rho_0^2-\frac{3\rho_0}{2\lambda}+\frac{1}{2\lambda^2}-\frac{ D_\rho \nu_r}{\lambda^2 v_0^2A^2}<0
\end{equation*}
solving this quadratic equation for $\rho_0$,
\begin{equation*}
    \rho_{0\pm}=\frac{3}{4\lambda} \pm \frac{1}{4\lambda}\sqrt{1+\frac{16D_\rho \nu_r}{v_0^2A^2}}
\end{equation*}

Rewriting the equations when the dimensionless activity $\frac{v_0^2}{D_{\rho} \nu_r}$ is larger and rearranging the terms we can write,

\begin{equation}
    \rho_{0} v_0^2 = A_1 v_0^2-\frac{B_1}{A^2}
    \label{eq10}
\end{equation}

\twocolumngrid
Here, $A_1=\frac{1}{3\lambda},B_1= \frac{2}{3\lambda}$ are parameter ($\lambda$) dependent constants. We have taken the values of fitting parameters $A_1 = 0.65, 0.45$  and $B_1 = 0.65, 0.45$ for $\rho_d = 0$ and $0.001$ respectively. We analyze the solution in Eq. \ref{eq10} and in Fig. \ref{fig3}(b), we have shown the straight green line plot between $\rho_0 v_0^2$  {\em vs.} $v_0^2$ for $\rho_d = 0$. The region above the straight line drawn with solid green line shows the instability or the phase separation for a clean system. However, in the presence of disorder $A = (1-\rho_d)$, ($\rho_d=0.001$) the straight dotted magenta line between $\rho v_0^2$ {\em vs.} $v_0^2$ shifts down and additional region marked with shaded color (orange) shows the instability. Which makes the mean density $\rho_0$ and activity $v_0$ shifts towards the smaller values for phase separation in the presence of weak disorder. 
 This observation is consistent with our numerical simulations, where the points in Fig. \ref{fig3}(b) are obtained from simulation data. 
The above diagram suggests that  disorder facilitates phase separation at reduced mean density ($\rho_0$) and particle activity ($v_0$).
 \\
We further investigate the transition from homogeneous to phase separated state for different disorder densities and activities. In Fig. \ref{fig3}(c), we show the plot of phase separation order parameter $\Delta \rho$ {\em vs.} $v_0$ for different: $\rho_d = 0.0$ - $0.2$ by fixing $\rho_0 = 0.7$. 
We observe a sudden jump in the $\Delta \rho$ at  $v_0 \sim 5.0$ for the clean system. In contrast to that, when we increase the disorder density slowly, the $\Delta \rho$ continuously changes from zero to finite values. Accordingly, the disorder makes the transition smooth  as found in previous studies \cite{villa2014quenched}.  In support of this finding, we calculated the effective free energy functional using linearized approximation. We plot the free energy functional $f(\delta\rho)$ $vs.$ $\delta\rho$ for clean as well as for $\rho_d= 0.2 - 1.0$ in Fig. \ref{free_en} in \ref{energy} showing the degree of asymmetry decreases with rise in the disorder density.  That indicates the transition from a homogeneous state to phase separated state becomes continuous type on increasing disorder density. The details of the analytical calculation can be found in Appendix \ref{energy}.


So far, we discussed the role of disorder on the steady state of the system. Next, it will be interesting to understand how the presence of disorder affects the kinetic of phase separation and morphology of domain walls. To do so, we focus on the parameter space where the homogeneous state is inherently unstable and undergoes phase separation.

{\em{Kinetics of phase separation}}:\\

It would be insightful to examine the growth law for domains in a disordered system compared to a clean system. The snapshots of local density field evolution for systems with $\rho_d$ = 0-0.5 are shown in Appendix \ref{time_evln} along with the density-density correlation functions (see Appendix \ref{main_cor}). 
 
Next, to illustrate the impact of disorder in the growth law, in Fig. \ref{fig6}(a)  we present a log-log plot of $L(t)$ {\em  vs.} $t$ for systems with varying levels of disorder $\rho_d$ from $0.0$ to $0.5$. 
The plot clearly shows that, asymptotically, the slope of $L(t)$ for $ \rho_d = 0.0, 0.05, 0.1, 0.2$ is approximately $0.33$, consistent with the usual Lifshitz-Slyozov (LS) growth law \cite{bray2002theory,puri2004kinetics} suggesting that phase separation in clean and low-disorder systems follow the standard LS law. Now, we show the nature of phase separation for low and moderate disorder densities with the help of two movies for disorder densities $\rho_d = 0.05$ and $\rho_d = 0.4$ respectively. The movie \href{https://drive.google.com/file/d/1ldM5_gEXsN5rTyixLtxnr7ChAzhuOq_S/view?usp=sharing}{MV2} shows the late time evolution of the local density of active particles around the pinning sites for $\rho_d = 0.05$. From the movie, it is observed that though initially the pinning sites act as nucleation center and promotes the phase separation by enhancing local clustering of the active density around them as shown in  \href{https://drive.google.com/file/d/1DFcUoYSQ5x2F0yMM5zccSz8v0i4wCeF0/view?usp=sharing}{MV1}, at sufficiently long time the mechanism of $MIPS$ dominates and large clusters (marked as (1) in MV2) wins at the cost of small clusters (marked as (2) in MV2). And finally, after very long time, all the small clusters are merged to one single big cluster (see \href{https://drive.google.com/file/d/1hVqsiiGm1wze4LUw4ZP5GYi8FnY5W0gA/view?usp=sharing}{MV3}).
That indicates the appearance of Macro phase separation for the systems having weak disorder densities.  However, as the disorder increases further i.e, for $\rho_d = 0.3, 0.4$, the size of the domains decrease and saturate at late times. That suggests the appearance of Micro phase separation at these disorder densities. The presence of moderate disorder sites, hinders the growth of Macro clusters and system goes to the Micro phase separation. To further characterise this Micro phase separation, we calculated the saturation length of $L(t)$ at late times $L_s$ and show its variation with respect to $\rho_d$ in Fig. \ref{fig6}(b) for three different system sizes. With increasing disorder density, it decays monotonically with $\rho_d$. We also calculated the saturation time $T_s$ for disorder densities moderate to high values $\rho_d = 0.3$ to $0.5$ for $K = 512$ (see inset plot). The presence of finite $T_s$ and $L_s$ confirms the Micro phase separation for intermediate disorder densities.\\
The movie \href{https://drive.google.com/file/d/1uLFiWfh1Spy1L2JUdFSlE1gSdqLd_V6I/view?usp=sharing}{MV4} shows there is a marginal growth around the disorder sites and its suppression illustrating the Micro-phase separation.  
Further, at very high disorder levels, such as $\rho_d = 0.5$, the plot shows no significant growth of $L(t)$, indicating a complete suppression of the usual phase separation kinetics. 

To interpret this more quantitatively,  we estimate the effective growth exponent $1/z_{eff}$
as a function of $t$ defines as $\frac{1}{z_{eff}} = \bigg<\frac{d lnL(t)}{d lnt}\bigg>$, where $<..>$ means average over late times. In Fig. \ref{fig6}(c), we showcase a semi-log $x$ plot depicting the variation of $\frac{1}{z_{eff}}$ with $\rho_d$. 
The plot represents that the dynamic growth exponent remains constant at 0.33 up to $\rho_d = 0.2$ showing that the LS law is followed. The vertical dashed line is the critical disorder density $\rho_d$ (obtained from the linearized hydrodynamic calculation as given in Appendix \ref{critical}, Eq. \ref{critical_rho}) \ after which we found Micro-phase separation and  at very high disorder density, there is no phase separation and only heterogeneous structures are formed. 

In Fig. \ref{fig6}(d), we show the scaled plot of $g(x)$ {\em vs.} scaled distance $x$ for different disorder densities, same as in Fig. \ref{fig5}. Very clearly for $\rho_d \le 0.1$, all the correlations show the nice collapse, indicating the static scaling for lower disorder densities, whereas the static scaling is not found for $\rho_d = 0.3, 0.5$ $(\rho_d > 0.2)$ .
\\
\begin{figure*}
    \includegraphics[width=1.0\textwidth]{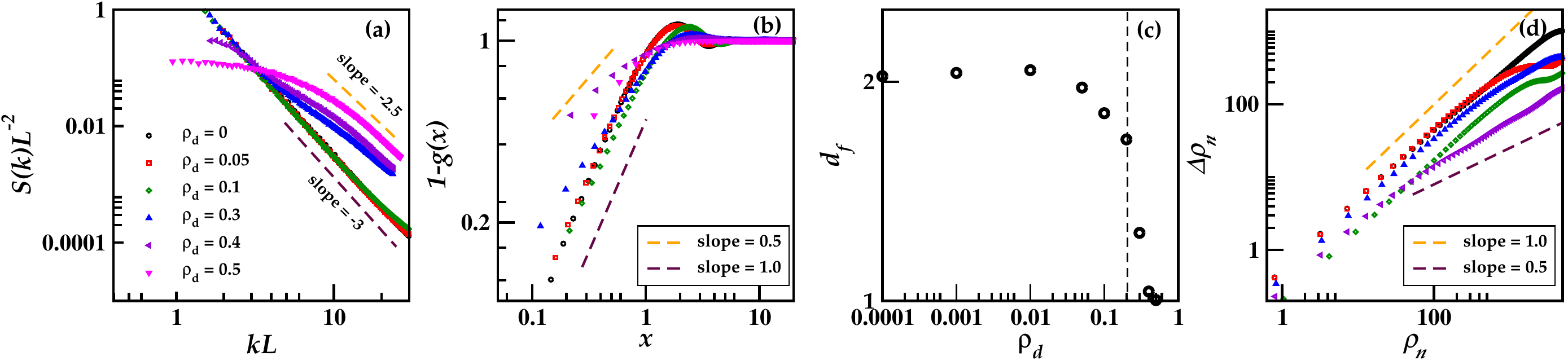}
    \caption{ The plot (a) exhibits the plot of scaled structure factor $S(k)L^{-2}$ {\em vs.} $kL$ for $\rho_d = 0- 0.5$ as shown in the legend. The maroon and orange dotted lines present the slope of line $-3$ and $-2.5$. The plot (b) displays $1-g(x)$ {\em vs.} $x$ for different disorder density  and the symbols have the same meaning as shown in (a). The legends present the slope of lines. (c) showcases the log-log plot of $d_f$ {\em vs.} $\rho_d$. The error bars are of the size of the symbols used. The vertical dashed line is drawn at $\rho_d = 0.23$ and represents the same meaning as the solid line in Fig. \ref{fig6}(c). The plot (d) depicts the density fluctuation $\Delta \rho_n$  {\em vs.} $\rho_n$ for different $\rho_d$. The symbols have the same meaning as in plot (a). The legends show the slope of lines as mentioned. }
    \label{fig7}
\end{figure*}
{\em Domain morphology and density fluctuations}:-\\
Till now, we have focused on the growth kinetics but, disorder also affects the domain morphology. 
To analyze the morphology in more details, including the interfacial properties, we calculate  the scaled structure factor $ S(k)L^{-d}$ and the cusp exponent $1-g(x) \sim x^{\theta} $. The exponent $\theta$ depends on the relevant morphology and reveals the roughness of the interface \cite{shrivastav2014non,banerjee2014fractal}. In Fig. \ref{fig7}(a), we present the plot of scaled structure factor $S(k)L^{-2}$  {\em vs.} $kL$ for system having $\rho_d$ from $0.0$ to $0.5$. Fig. \ref{fig7}(a) are already scaled with respect to time for individual system having different $\rho_d$. The range of time is the same as that for the $C(r)$ shown in Fig. \ref{fig5}(a-e). We found all the curves for $\rho_d \le 0.1$ shows the nice static scaling as well, however, static scaling is not observed for $\rho_d > 0.1$ (0.3, 0.4, 0.5) as shown in Fig. \ref{fig7}(a). It suggest  that the morphologies of the domain changes on varying disorder densities. 
For disorder densities up to $0.1$, we observe the Porod's tail $S(k)L^{-2} \sim (kL)^{-3}$ implying the sharp interface between the domains. However, above $\rho_d > 0.1$, the behavior deviates from Porod's tail, approaching $S(k)L^{-2} \sim (kL)^{-2.5}$ which suggests the emergence of indistinct boundary between domains. 
The roughening of interfaces in some non-equilibrium systems are also observed in some of the recent works \cite{kushwaha2025unconventional,PhysRevLett.130.187102,PhysRevLett.81.1469,shrivastav2014non,banerjee2014fractal}.
 We also compared the structure factor calculated from linearised hydrodynamic (see Appendix \ref{structure factor}) and it shows the structure factor flattens with increasing disorder the same as found in the numerically in Fig. \ref{fig7}(a).  \\
Further, to examine the roughness of the interface, in Fig. \ref{fig7}(b), we plot $1-g(x)$ {\em vs.} $x$ for different disorder densities ranging from $\rho_d = 0.0 - 0.5$. We observe a crossover of $\theta$ from 1 to 0.5.
For $\rho_d \le 0.1$, the morphology of the domains are smooth, whereas we notice the rough domains for $\rho_d > 0.1$ (0.3, 0.4, 0.5). We examine the fractal nature of the interfaces by calculating their fractal dimension using the box-counting method. The detailed method is provided in the appendix \ref{fractal}. We showcase a log-log plot of the fractal dimension of the interface $d_f$ as a function of disorder density $\rho_d$ in Fig. \ref{fig7}(c). The plot reveals a smooth crossover from $d_f =2$ in a clean system to $d_f = 1$ in a strongly disordered system, confirming the emergence of fractality and increased irregularity in the interface due to disorder.   \\

We also calculated the density fluctuations in the system $\Delta \rho_n$. The detail of the calculations is provided in the Appendix \ref{nf}. In Fig. \ref{fig7}(d), we show the plot of $\Delta \rho_n$ {\em vs.} $\rho_n$ for different $\rho_d = 0.0$, $0.05$ , 0.1, 0.3 and $0.5$. For $\rho_d \le 0.1$, the density fluctuations is large \cite{PhysRevLett.108.238001} and $\Delta \rho_n $ goes as $\rho_n$, whereas for large $\rho_d$, $\Delta \rho_n $ $\sim$ $\sqrt{\rho_n}$ for large $\rho_n$. That suggests the diffusive nature of the particles for large disorder in the system. 

 \section{Discussion}
Disorder plays an important role due to its inherent intrinsic and extrinsic presence in natural systems, making its study highly relevant. Recently, research on disordered active matter has gained significant attention, emerging as an important area of study. In this work, we proposed a coarse-grained model of a collection of self-propelled particles to explore the phase behavior in the presence of pinned disorder. We use the disorder as pinning sites where the velocity of the particles become zero when it comes in contact to it. The key finding of our work is, weaker level of  disorder promotes phase separation in the system within the phase space of $v_0$ and $\rho$ at lower values i.e. below the threshold values for $MIPS$ in a clean system. Additionally, the kinetics slows down at moderate disorder, and shows the Micro phase separation. The Micro phase separation is also found in some recent studies \cite{kushwaha2025unconventional,yadav2025coarsening,PhysRevX.8.031080}. 
However, beyond a certain threshold the system's kinetics become arrested and phase separation suppresses.\\

Along with these key results, we conduct a detailed investigation of disordered system. Our analytical calculation using linear stability analysis shows a good agreement with the numerically obtained phase boundary at lower disorder level. We report that transition from the non-phase separated state to the phase-separated state in the system exhibits a crossover from discontinuous to smooth behavior with respect to $v_0$ as the pinning sites in the system increases. 
By analyzing  the characteristic length scales and effective exponents, we quantify how the domain growth deviates from conventional LS law beyond a threshold of disorder density showing Micro phase separation and kinetic arrest at higher disorder. The tail of the structure factor fits well with Porod's tail for low disorder densities, but gradually deviates beyond a critical threshold, indicating increased interface roughness. This signature is observed in the cusp exponent which exhibits a crossover from 1 to 0.5 roughness of interface in the disordered system. Additionally, we calculate the fractal dimension of interface as a measure of it's irregularity and find that $d_f$ varies from 2 to 1 with disorder. Further, we report the density fluctuations become giant in clean as well as weak disordered system and suppressed in systems with strong disorder. 
All the results combining in a bigger frame shows how disorder affects the system's phase behavior, kinetics and dynamical properties. \\

The study of systems with disorder is very important  as it allows researchers to explore how complex, real-world environments influence the dynamics and statistics of active systems. Disorder can arise in various forms, such as spatial heterogeneity, temporal fluctuations, or randomness in particle properties.  Many natural systems, such as bacterial colonies, cell tissues, and animal groups, operate in environments that are inherently disordered. In this work, we provide a fundamental study on active matter system's phase behavior with spatial heterogeneity as pinned disorder and can be helpful for understanding the behavior of biological systems in presence of disorder. \\

In this study, we explore the system with quenched obstacles, in future it would be interesting to explore with diffusive obstacles.

\label{sec:discussion}
\section{Data availability}
The datasets used and/or analysed during the current study available from the corresponding author on reasonable request.
\begin{acknowledgments}
P.J. thanks Prasenjit Das for giving useful comment on the work. P.J. gratefully acknowledge the DST INSPIRE fellowship  for
funding this project and S.D thanks the IIT (BHU) for funding this project. The support and the resources provided by PARAM Shivay Facility under the National Supercomputing Mission, Government of India at the Indian Institute of Technology, Varanasi are gratefully acknowledged by all authors. S.M. thanks DST-SERB India, ECR/2017/000659, CRG/2021/006945 and MTR/2021/000438  for financial support. P.J. and S.M. also thank the Centre for Computing and Information Services at IIT (BHU), Varanasi.
\end{acknowledgments}
\onecolumngrid
\appendix
\renewcommand{\thefigure}{A\arabic{figure}}
\setcounter{figure}{0} 
\section{The correlation between pinned sites and particle density $C_{\rho\rho_d}(t)$}\label{cor_t}
\begin{figure} 
     \centering
     \includegraphics[width=0.4\textwidth]{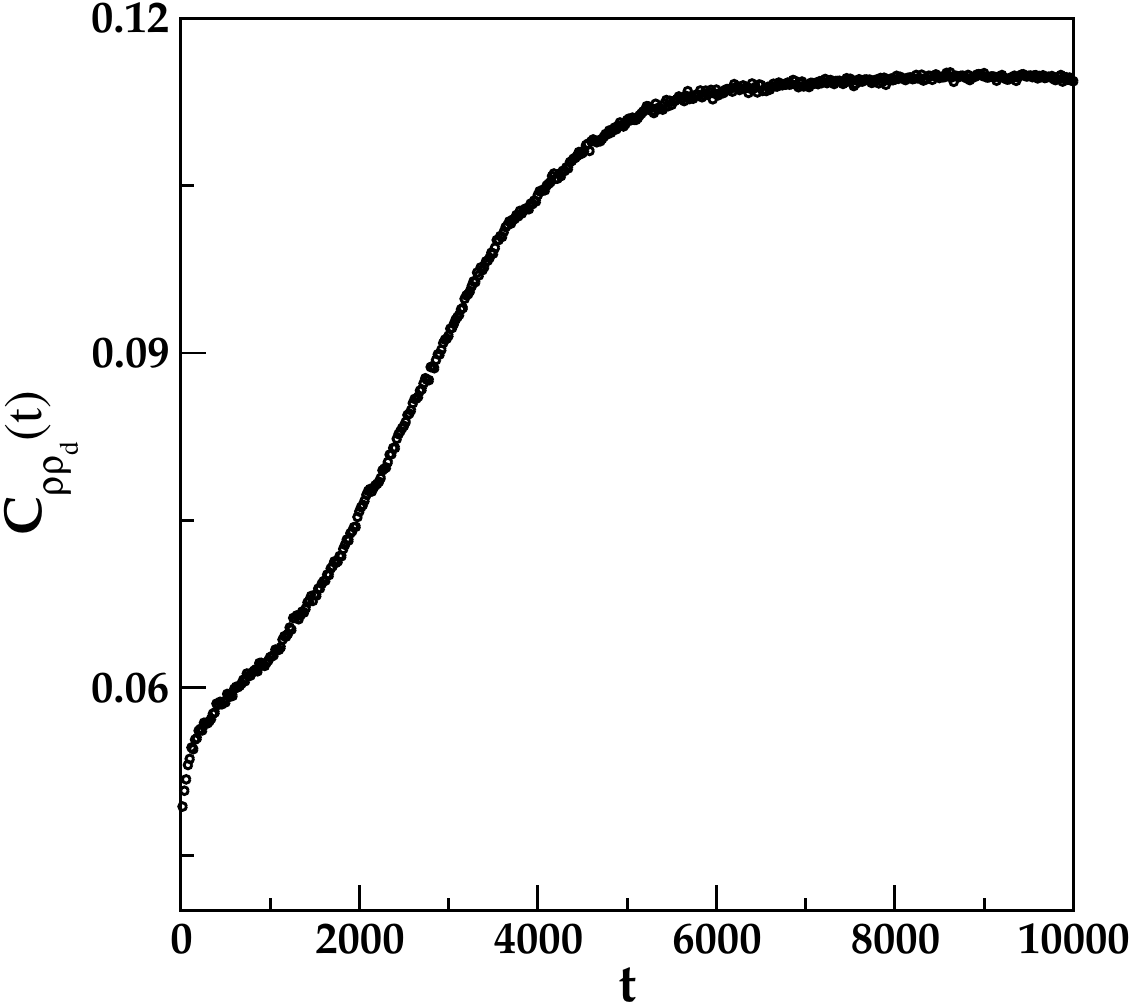}
     \caption{The plot showcases the $C_{\rho \rho_d}$ $vs.$  $t$ for $\rho_d = 0.05$. The plot is generated for system size $K = 512.$}
    \label{Fig:A2}
 \end{figure}

To show the mechanism of disorder induced phase separation $(DIPS)$ quantitatively, we calculate the correlation defined as $C_{\rho \rho_d}(t) = \frac{1}{\bar K^2} \sum_{ij} (\delta\rho_d(r_i) \times \delta \rho(r_i))$.
The $\delta\rho_d(r_i)$  and $\delta \rho(r_i)$ are the fluctuations of disorder density and density  of active particles respectively (in binary representation) from their mean values. Hence regions above/below the mean density are marked as $+1$ and $-1$ respectively.
Here, we take the mean density of the system $\rho_0 = 0.6$ and disorder density $\rho_d=0.05$. To calculate the fluctuations we coarse-grained the box in to $8\times 8$ sizes. $r_i$ is the coordinate on the lattice obtained after coarse-graining, and normalization $\bar{K}^2$ is number of coarse-grained cells. Fig. \ref{Fig:A2} shows the time series of correlation for $\rho_d=0.05$.

 \section{Time evolution of local density field for the systems is different values of disorder}\label{time_evln}
\begin{figure*}
    \centering
    \includegraphics[scale=0.36]{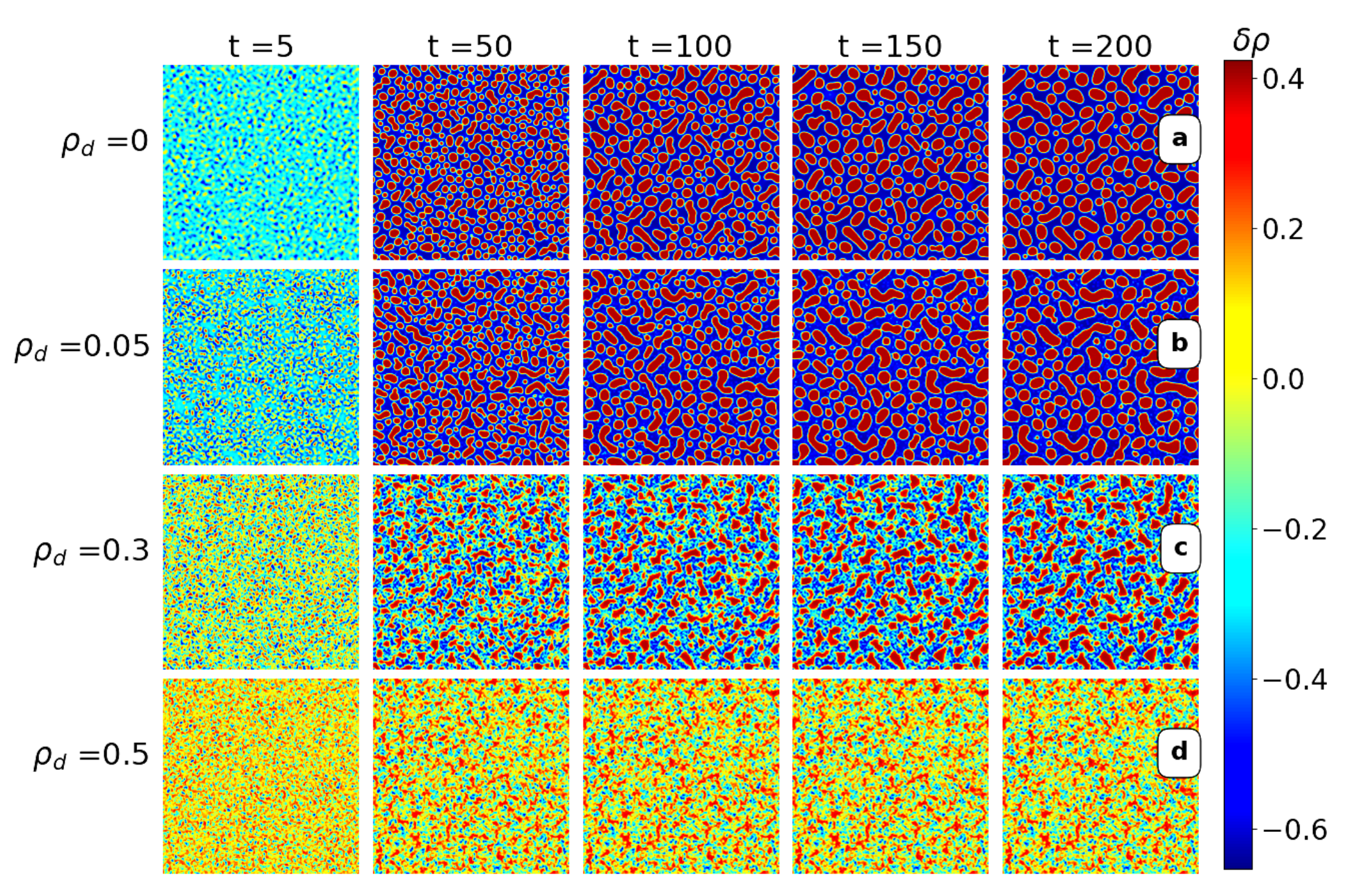}
    \caption{The panels (a-d) showcase the snapshots of time series of the local density field $\rho$ at t = 5, 50, 150 and 200 across the columns in sequence. Each panel from top to bottom, depicts the snapshots for system with $\rho_d = 0, 0.05, 0.3, 0.5$, respectively. The color in the heatmap represent the magnitude of local density at each lattice point. The results are obtained from simulating a $256 \times 256$ system. }
    \label{fig4}
\end{figure*}

 Fig. \ref{fig4} (a-d) depict the snapshots of the local density fluctuations $\delta\rho(\boldsymbol{r},t)$ for $\rho_d = 0 - 0.5$ in sequence at different times $t = 5 - 200$ as shown in each row of (a-d). 
 From the snapshots, it can be seen that the density inhomogeneity is observed for $\rho_d = 0, 0.05, 0.3$ in Fig. \ref{fig4}(a-c). However, in case of higher disorder i.e, $\rho_d = 0.5$ , the density homogeneity is suppressed showing almost no phase separation in Fig. \ref{fig4}(d). 
\section{Density-density correlation functions}\label{main_cor}
\begin{figure*}
     \includegraphics[scale=0.43]{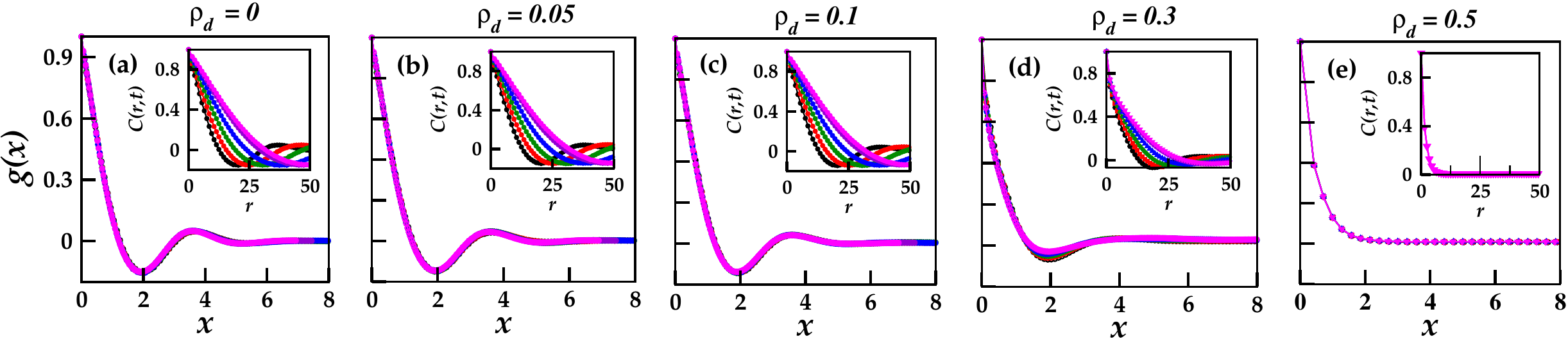}
    \includegraphics[scale=0.38]{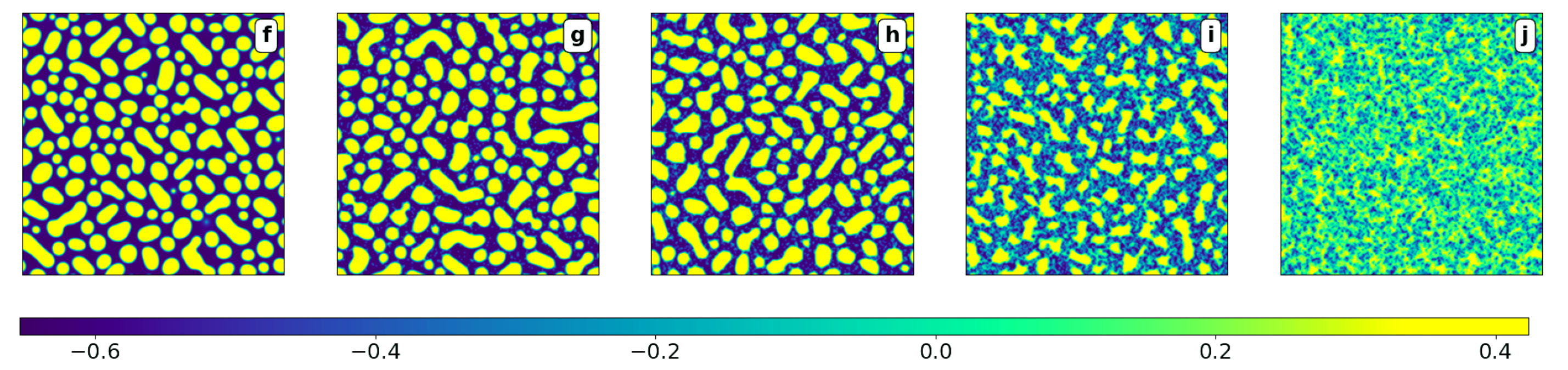}
   
    \caption{ 
    The main plot (a-e) present the scaled $g(x) $ {\em vs.} $x$ for disorder density $ \rho_d = 0-0.5$ respectively. The insets of each plot shows time progression ($t = 1000-20000$) of correlations by plotting the $C(r,t) $ {\em vs.} $r$ for corresponding disorder density. The correlation functions are obtained from system of $K= 512$ averaging over 50 ensembles. The snapshots (f-j) showcase the local density fluctuation $\delta\rho$ at $t = 1000$ for $\rho_d = 0.0, 0.05, 0.1, 0.3, 0.5$. The heatmap shows the magnitude of $\delta\rho$ at each lattice point. We have taken a $256 \times 256$ system to obtain the snapshots. }
    \label{fig5}
\end{figure*}

  A standard tool to obtain information about sizes and
textures of domains and interfaces is the two-point spatial
correlation function of density fluctuations defines as $C(r,t)$ = $\langle \delta\rho({\bf r}'+{\bf r},t) \delta\rho({\bf r}',t)   \rangle$ and the corresponding Fourier transform i.e, structure factor $S(k, t) =  \langle \delta \rho({\bf k}'+{\bf k},t) \delta \rho({\bf k}',t)   \rangle$.
 The $\langle....\rangle$	 denotes an average over reference positions ${\bf r}'$, and $50$ independent realizations.
In Fig. \ref{fig5}(a-e), we present the correlation functions for systems having disorder densities ranging from $\rho_d = 0 - 0.5$ respectively. The insets in each plot depict the progression in time of the correlation function. The decay of correlations is slower with time, suggesting an increase in the size of clusters. The time evolution of the correlations are similar in the system with $\rho_d = 0, 0.05, 0.1$ in Fig. \ref{fig5}(a-c) (inset), while a slower increase in correlations observed in system with $\rho_d = 0.3$ in Fig. \ref{fig5}(d) (inset). Further, in case of high disorder density $\rho_d = 0.5$, the temporal development of correlation is arrested showing sharp decay in correlations as shown in Fig. \ref{fig5}(e) (inset). 

The main plots of Fig. \ref{fig5}(a-e) display the scaled correlation functions obtained as $g(x)$ , where $x= r/L(t)$. 
Here, $L(t)$ is the characteristic length determined as the distance over which the correlations cross 0.1.
The correlation functions show a dynamical scaling showing a good scaling collapse across all disorder cases. That implies the existence of a single characteristic length scale $L(t)$ and the evolution morphology can be characterized by a distinct single length scale for each disorder level. Although, the weak disorder promotes phase separation, the decrement in the two-point correlation functions in the disordered system, make it  distinct from the $MIPS$. 
In Fig. \ref{fig5}(f-j), we show the snapshots for local density fluctuation $\delta\rho(\boldsymbol{r})$ in the system having the corresponding values of $\rho_d = 0- 0.5$ respectively at time $t = 1000$. In Fig. \ref{fig5}(f-h), we observe domain formation showing stronger density contrast up to $\rho_d = 0.1$, while in the case of $\rho_d = 0.3$, phase separation is observed with smaller density inhomogeneity as shown in Fig. \ref{fig5}(i) and  there is formation of heterogeneous structures and no  clear phase separation for $\rho_d = 0.5$ as shown in Fig. \ref{fig5}(j). The similar results are obtained in previous study of \cite{ro2021disorder} for disordered system. These snapshots justify the above correlation plots nicely.
\section{Linearised calculations}

\subsection{{Calculation of Structure factor}}

From the Eq. \ref{A7}, the solution for $\delta \rho$ is;

\begin{equation}
  \delta \rho(\boldsymbol{q},\omega) =   \frac{[-(-i \omega+kq^2+\nu_r)i qf_q-iqf_PV]}{(\omega-i \omega_1)(\omega-i \omega_2)}
\end{equation}
The correlation function is defined as  $C_{\rho\rho}(\boldsymbol{q},\omega)$ = $\langle \delta \rho(\boldsymbol{q}, \omega) \delta \rho(\boldsymbol{-q}, -\omega)\rangle$;

\begin{equation}
    \langle \delta \rho(\boldsymbol{q},\omega)  \delta \rho(\boldsymbol{-q}, -\omega)\rangle = \frac{[\omega^2+(kq^2+\nu_r)^2]q^2\Delta_{\rho}+V^2q^2\Delta_P }{(\omega^2+\omega_1^{2})(\omega^2+\omega_2^{2})}
\end{equation}
The static structure factor is defined as,  \[S(\boldsymbol{q}) = \int_{-\infty}^{\infty} C_{\rho\rho}(\boldsymbol{q},t) \,d\omega \]
 The full expression for the static structure factor is;

\begin{equation}
     S(q) = \frac{q^2\Delta_\rho\pi}{(D_q+k)q^2+\nu_r}+\frac{\pi v_0^{2}(1-\lambda\rho_0)^2A^2\Delta_P}{4[(kq^2+\nu_r)D_\rho+\frac{A^2}{2}v_0^{2}(1-\lambda\rho_0)(1-2\rho_0\lambda)][(D_\rho+k)q^2+\nu_r]}
     \label{s_dis}
\end{equation}
For clean system limit, setting $A = 1$ we can get,
\begin{equation}
    S(q) = \frac{q^2\Delta_\rho\pi}{(D_q+k)q^2+\nu_r}+ \frac{\pi v_0^{2}(1-\lambda\rho_0)^2\Delta_P }{4[(kq^2+\nu_r)D_\rho+ \frac{v_0^2}{2}(1-\lambda\rho_0)(1-2\rho_0\lambda)]+(D_\rho+k)q^2+\nu_r)}
\end{equation}
which matches well with the result obtained in \cite{fily2012athermal}. In the presence of disorder ($A = 1-\rho_d$), the  second term in eq. \ref{s_dis} flatten the structure factor. This particular behavior is also observed in Fig. \ref{fig7}(a), where for clean system and weak disorder, the decay is sharp but for higher $\rho_d$ the plot becomes flatter and the tail of structure factor deviates from Porod's tail. 
\subsection{Determination of critical $\rho_d$ from the structure factor}\label{critical}
To obtain the critical value of $\rho_d$ after which the phase separation suppresses, we observe at which critical value of $\rho_d$ the denominator of the structure factor if positive and has a finite value for small wave vector limit.
We rewrite the structure factor in the $q \xrightarrow{}0$ limit;
\begin{equation*}
     S(q) = \frac{\pi v_0^{2}(1-\lambda\rho_0)^2A^2\Delta_P }{4\nu_r[\nu_rD_\rho+\frac{A^2}{2}v_0^{2}(1-\lambda\rho_0)(1-2\rho_0\lambda)]}
      = \frac{\bar\Delta}{D_{eff}}
\end{equation*}

We define $D_{eff} = \nu_rD_\rho+\frac{A^2}{2}v_0^{2}(1-\lambda\rho_0)(1-2\rho_0\lambda) $ and $|(1-\lambda\rho_0)(1-2\rho_0\lambda)|=\gamma(\rho_0)$. Here, $\bar\Delta = \pi v_0^{2}(1-\lambda\rho_0)^2A^2\Delta_P/(4 \nu_r)$. 
After a critical disorder density $\rho_d$, the $D_{eff}$ become positive and this implies
\begin{equation*}
    \nu_rD_\rho > \frac{(1-\rho_d)^2}{2} v_0^2\gamma(\rho_0)
    \end{equation*}
    We can write, 
\begin{equation}
         \rho_d(\rho_0, v_0)> 1-\frac{1}{v_0}\sqrt{\frac{2D_{\rho}\nu_r}{\gamma(\rho_0)}}
 \label{critical_rho}
\end{equation}
    
    Substituting the values of parameters used in the numerical simulation, $\rho_0 = 0.7$ and $v_0 = 6.0$, we obtain the critical value of $\rho_d \geq 0.23$. 
    \label{structure factor}
\subsection{Calculation of effective free energy}\label{energy}
 In this part we derive the free energy functional $F{{\delta \rho(r)}}$. 
 Starting from the {\bf P} equation, neglecting diffusion,
\begin{equation}
   \partial_{t}\boldsymbol{P}=-\nu_{r} \boldsymbol{P}-\frac{1}{2}\boldsymbol{\nabla}(v_{0}(1-\lambda\rho)A(r)\rho)+\boldsymbol{f_{P}}
   \label{eq:v1}
\end{equation}
In the steady state, $$\delta_{t}\boldsymbol{P}=0 $$. 
\begin{equation*}
    \boldsymbol{P}=-\frac{1}{2\nu_{r}}\boldsymbol{\nabla}((v_{0}(1-\lambda\rho)A(r))
\end{equation*}

Adding small perturbation $\delta\rho$, the equation becomes,
\begin{equation*}
 \boldsymbol{P}=-\frac{1}{2\nu_{r}}\boldsymbol{\nabla}[v_{0}(1-\lambda(\rho_{0}+\delta\rho))A(r)(\rho_{0}+\delta\rho)]
\end{equation*}

\begin{equation*}
 \boldsymbol{P}=-\frac{1}{2\nu_{r}\rho_0}\boldsymbol{\nabla}[(v_{e}-v_{0}\lambda\delta\rho)A(r)(\rho_{0}+\delta\rho)]
\end{equation*}
here $v_{e}=v_0(1-\lambda\rho_0)$
\begin{equation*}
 \boldsymbol{P}=-\frac{v_0}{2\nu_{r}\rho_0}\boldsymbol{\nabla}[A(r)({\delta\rho(1-2\lambda\rho_0)-\lambda\delta\rho^2+\rho_{0}(1-\lambda\rho_{0}))}]
\end{equation*}
substituting the value of $\delta \boldsymbol{P}$ in $\delta \rho$ equation ,

 \begin{equation}
    \partial_{t}\rho= -\boldsymbol{\nabla}.(v_0(1-\lambda\rho)A(r)\boldsymbol{P}-D_{\rho}\boldsymbol{\nabla}\rho)
    \end{equation}
\begin{equation*}
  \partial_{t}\rho=-\boldsymbol{\nabla}.[(v_{e}-v_{0}\lambda\delta\rho)A(r)]\boldsymbol{\delta P}+D_\rho\nabla^2 \delta\rho
\end{equation*}

\begin{equation*}
  \partial_{t}\rho=\boldsymbol{\nabla}.[(v_{e}-v_{0}\lambda\delta\rho)A(r)][\frac{v_0}{2\rho_0\nu_{r}}\boldsymbol{\nabla}[A(r)({\delta\rho(1-2\lambda\rho_0)-\lambda\delta\rho^2+\rho_{0}(1-\lambda\rho_{0}))}]]+D_{\rho}\boldsymbol{\nabla}^2\delta\rho
\end{equation*}

\begin{equation*}
    \partial_{t}\rho=\frac{v_0}{2\rho_0\nu_{r}}\boldsymbol{\nabla}.[(v_e-v_0\lambda\delta\rho)A(r)\boldsymbol{\nabla}(A(r)(\delta\rho(1-2\lambda\rho_0)-\lambda\delta\rho^2+\rho_0(1-\lambda\rho_0)]+D_\rho\boldsymbol{\nabla}^2\delta\rho
\end{equation*}
 
\begin{equation*}
    \partial_{t}\rho=-\boldsymbol{\nabla}.\boldsymbol{J}
\end{equation*}
$\boldsymbol{J}$ can be written as,
\begin{equation*}
    \boldsymbol{J}=\frac{v_0}{2\rho_0\nu_{r}}[(v_e-v_0\lambda\delta\rho)A(r)\boldsymbol{\nabla}(A(r)(\delta\rho(1-2\lambda\rho_0)-\lambda\delta\rho^2+\rho_0(1-\lambda\rho_0)]+D\boldsymbol{\nabla}\delta\rho
\end{equation*}

$\boldsymbol{J}$ is also related to the chemical potential,
\begin{equation*}
    \boldsymbol{J}=-\boldsymbol{\nabla}\mu
\end{equation*}

The chemical potential is related to free energy by,
\begin{equation*}
    \mu=\frac{v_0}{4\rho_0 \nu_r}( A(r)(\rho_0(1-\lambda\rho_0-\lambda\delta\rho^2-2\lambda\delta\rho+D\delta\rho+\delta\rho)))^2
\end{equation*}
\begin{equation*} 
    \mu=\frac{\delta f(\delta\rho)}{\delta(\delta\rho(r))}
\end{equation*}
free energy can be obtained by integrating $\mu$ over $\delta\rho$ ,

\begin{equation*}
    f(\delta\rho)=\frac{v_0}{4\rho_0 \nu_r}\int d^dr\int (A(r) v_0 \rho_0 (1-\lambda\rho_0-\lambda\delta\rho^2/\rho_0-2\lambda\delta\rho+\delta\rho/\rho_0+D\delta\rho))^2 d\delta\rho(r)
\end{equation*}

Integrating w.r.t $\delta\rho$ first,
\begin{equation*}
     f(\delta\rho)= \frac{v_0 A(r)}{4\rho_0 \nu_r}[\rho_0^2(1-\lambda\rho_0)^2\delta\rho+(1-\lambda\rho_0)(1-2\lambda\rho_0)\rho_0 \delta\rho^2+(\frac{1}{3}(1-2\lambda\rho_0)-\frac{2}{3}(1-\lambda\rho_0)\lambda\rho_0)\delta\rho^3 - \frac{\lambda}{2}(1-2\lambda\rho_0)\delta\rho^4+\frac{\lambda}{5}\delta\rho^5]
\end{equation*}

\begin{equation*}
     f(\delta\rho)=A(r)(A_{11}\delta\rho+A_{12}{\delta\rho^2}+{A_{13}\delta\rho^3}+{A_{14}\delta\rho^4}+{A_{15}\delta\rho^5})
\end{equation*}

\begin{figure}
    \centering
    \includegraphics[width=0.4\linewidth]{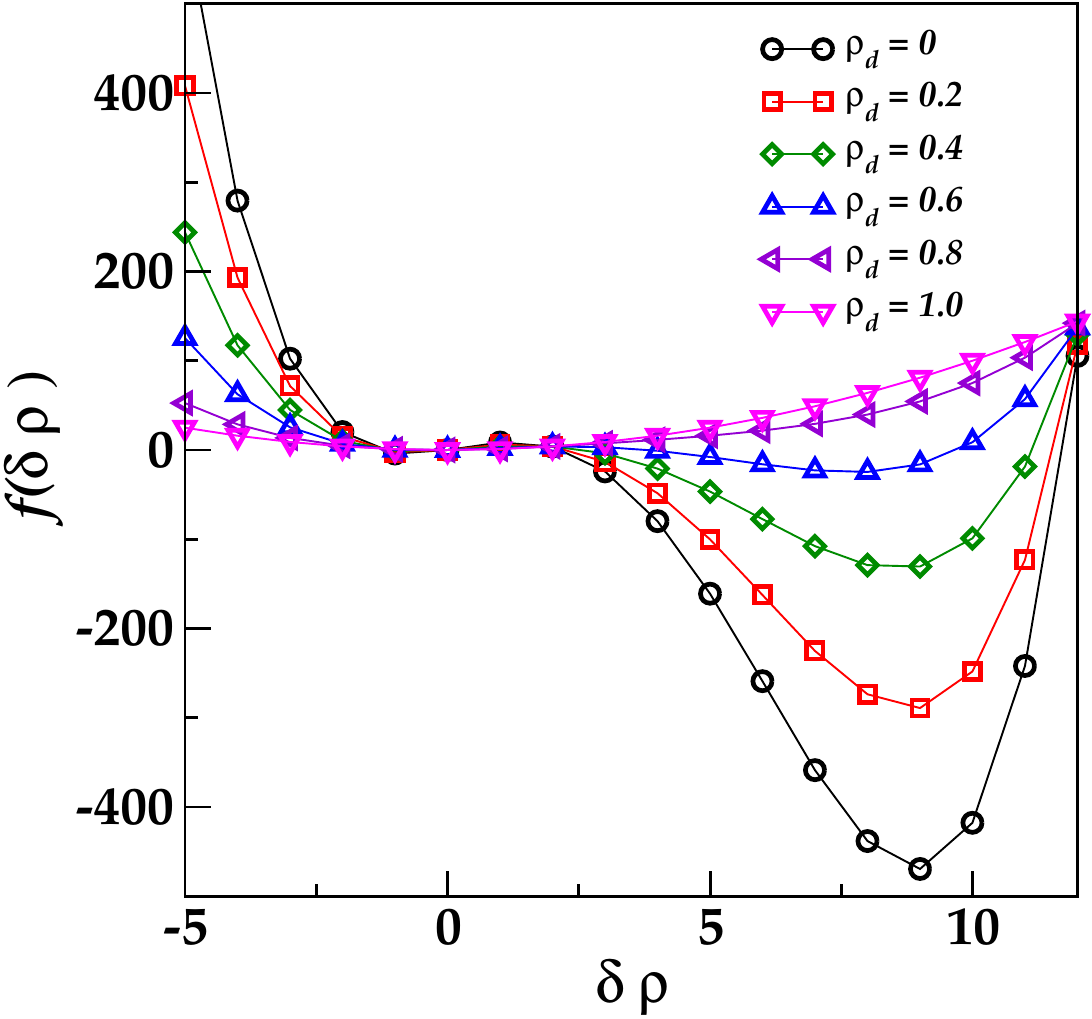}
    \caption{The plot shows the free energy functional $f(\delta\rho)$ {\em {\em {\em vs.}}} $\delta\rho$ for different $\rho_d$ as shown by the legends. }
    \label{free_en}
\end{figure}
In Fig. \ref{free_en}, we plot the free energy functional $f(\delta\rho)$ {\em {\em {\em vs.}}} $\delta\rho$ for different $\rho_d$. The numerical values of all the coefficients of $\Delta \rho$ is taken from the parameters used in the numerical study. From the Fig, it is observed that with rise of the disorder, the degree of asymmetry in the free energy functional  decreases and the it becomes symmetric for higher disorder densities. This is the indication of the transition from homogeneous to unstable phase becomes continuous or second order transition. 

\section{Fractal dimension of interface} \label{fractal}
To investigate the fractal nature of the domains we plot the fractal dimension of the interface $d_f$ {\em {\em {\em vs.}}} disorder $\rho_d$ in Fig.7 (c) (main manuscript). To calculate the fractal dimension of the interface we follows the below procedure: For a fractal interface, the length of the interface decreases with time $t$ as $B(t) \approx t^{-\frac{d_f - 1}{3}}$, where, $d_f$ is the fractal dimension \cite{streitenberger2000coarsening}. 
The length of the interface is determined using a box-counting algorithm. We begin by evaluating the fluctuation of the order parameter at each lattice point $(i,j)$ as $\delta\rho(i,j) = \rho(i,j)-\rho_0$ where, $\rho_0$ is the mean value of $\rho$. A point is classified as part of a high-density region if $\delta\rho >0$, otherwise, it is considered to belong to a low-density region. Each lattice point is then further categorized based on the number of its high-density neighbors, denoted by by $n_h$. Points with $n_h = 4$ are considered core points, residing within clusters. Points with $2\le n_h \le 3$ that has at least one neighbor which is a core point are classified as edge points, lying at the interface. Points outside these criteria are deemed isolated, located in low-density regions.
In this study, the spatial grid size is uniform in both 
$x$ and $y$-directions. The perimeter or length of the interface, $B(t)$, is then calculated by multiplying the number of edge points by the spatial grid size. 

\section{Density fluctuation}\label{nf}
To calculate the density fluctuations in the system $\Delta \rho_n = \sqrt{\langle\rho^2\rangle-\langle\rho\rangle^2}$, we divide the simulation box of size $512\times 512$ into square blocks of different sizes from $1 \times1$ to $256 \times 256$. For each block size, we calculate the total density within each block across multiple time steps in the steady state and  determine the average density $\rho_n$, so as the fluctuations from that average. By repeating this process for different block sizes, we calculate average densities $\rho_n$ and their variances $\Delta \rho_n$ for different system with $\rho_d = 0 - 0.5$
\section{Description of Movies}
{\em Movie1(MV1)-} In MV1, we show the early time evolution of the local density field $\rho(r)$ of the active particles for system size $K=128$ and disorder density $\rho_d = 0.05$. The real time is displayed on the top of each snapshots. The black dots represents the pinning sites or disorder sites. The color bar presents the magnitude of the local density field at the lattice points.\\
{\em Movie2(MV2)-} In MV2, we illustrates the late time evolution of the local density field $\rho(r)$ for same system (Real time is mentioned on the top). And the black dots and color bar represents the same as in MV1. The first two prominent clusters are identified using the  numbers (1-2) in the movie.\\
{\em Movie3(MV3)-} The MV3, shows the very late time evolution of the local density field $\rho(r)$ for same system (Real time is mentioned on the top).\\
{\em Movie4(MV4)-} The MV4, showcases the time evolution of the local density field $\rho(r)$ for system having $\rho_d= 0.4$ (Real time is mentioned on the top). The other details are  the same as for the above MV1.

\twocolumngrid 
\nocite{*}

%
\end{document}